\documentclass[aps,pre,notitlepage,twocolumn,superscriptaddress]{revtex4-1}

\usepackage{amsmath,amsfonts}
\usepackage{mathrsfs}
\usepackage{graphicx,color}
\usepackage{bm,bbm}
\usepackage{multirow}
\usepackage[export]{adjustbox}
\usepackage{hyperref}
\usepackage[caption=false]{subfig}
\usepackage{algorithm,algpseudocode}
\usepackage{longtable}

\def\OO{\mathcal O}

\begin{document}

\title{Percolation thresholds on high dimensional $D_n$ and dense packing lattices}
\author{Yi Hu}
\affiliation{Department of Chemistry, Duke University, Durham, North Carolina 27708, USA}
\author{Patrick Charbonneau}
\affiliation{Department of Chemistry, Duke University, Durham, North Carolina 27708, USA}
\affiliation{Department of Physics, Duke University, Durham, North Carolina 27708, USA}

\date{\today}

\begin{abstract}
The site and bond percolation problems are conventionally studied on (hyper)cubic lattices, which afford straightforward numerical treatments. 
The recent implementation of efficient simulation algorithms for high-dimensional systems now also facilitates the study of $D_n$ root lattices in $n$ dimension as well as $E_8$-related dense packing lattices. 
Here, we consider the percolation problem on  $D_n$ for $n=3$ to $13$  and on $E_8$ relatives for $n=6$ to 9. 
Precise estimates for both site and bond percolation thresholds obtained from invasion percolation simulations are compared with dimensional series expansion on $D_n$ lattices based on lattice animal enumeration. 
As expected, the bond percolation threshold rapidly approaches the Bethe lattice limit as $n$ increases for these high-connectivity lattices. 
Corrections, however, exhibit clear yet unexplained trends. Interestingly, the finite-size scaling exponent for invasion percolation is found to be lattice and percolation-type specific. 
\end{abstract}

\maketitle

\section{Introduction}

Percolation being one of the simplest critical phenomena, its models play particularly important roles in statistical physics \cite{stauffer1994percolation}. Minimal models -- lattice-based ones, in particular -- have thus long been used to test notions of universality as well as the relationship between mean-field and renormalization group predictions. 
On lattices, two covering fractions $p$ can be defined: (i) the probability that a vertex is occupied, and (ii) the probability that an edge between nearest-neighbor vertices is occupied. As $p$ increases, a percolating cluster forms at a threshold $p_\mathrm{c}^\mathrm{site}$ or $p_\mathrm{c}^\mathrm{bond}$, depending on the covering choice~\cite{stauffer1994percolation}. 
Because precise thresholds values are prerequisite for stringently assessing criticality~\cite{mertens2018percolation,huang2018critical,biroli2019dynamics,xun2020precise},
(and because thresholds are lattice specific and lack an analytical expression~\cite{wierman2002accuracy},) substantial efforts have been directed at estimating them through numerical simulations~\cite{lorenz1998precise,xu2014simultaneous,mertens2018percolation,kotwica2019efficient,xun2020bond} and graph-based polynomial methods~\cite{scullard2008critical,scullard2010critical,jacobsen2015critical,scullard2020bond}. The strong dependence of criticality on spatial dimension $n$, especially above and below its upper critical dimension, $n_\mathrm{u}=6$, motivates expanding these efforts over an extended range of $n$~\cite{kirkpatrick1976percolation,stauffer1994percolation}. 

In this context, the invasion percolation algorithm recently introduced by Mertens and Moore~\cite{mertens2017percolation,mertens2018percolation} is particularly interesting. In short, the algorithm directly grows a percolating cluster, and thus provides both the universal asymptotic critical behavior and the lattice-specific finite-size scaling correction. Most crucially, by avoiding the explicit construction of a lattice grid, the scheme preserves a polynomial space complexity as $n$ increases. Threshold values up to ten significant digits of precision have thus been obtained on hypercubic lattices ($\mathbb{Z}^n$) up to $n=13$~\cite{mertens2018series}. 

Hypercubic lattices, although geometrically straightforward, are in some ways not natural systems to consider as dimension increases. Recall that lattices can be seen as discretizations of Euclidean space $\mathbb{R}^n$ in which each lattice point is a vertex of a cell in that tessellation. As $n$ increases, the cubic cells that tile $\mathbb{Z}^n$ become increasingly dominated by spiky corner sites.
The cells of root lattices, $D_n$, are relatively less spiky in $n \ge 3$. A way to quantify this effect is to compare the sphere packing fraction for different lattices. In this sense, $D_n$ packings  are exponentially denser than their $\mathbb{Z}^n$ counterpart by a factor of $2^{n/2-1}$~\cite{Conway1988}.
Similarly, the eight-dimensional $E_8$ lattice corresponds to a sphere packing fraction twice that of $D_8$ (16 times that of $\mathbb{Z}^8$); $E_8$-related lattices, $E_6$, $E_7$ and $\Lambda_9$ are also the densest known sphere packings in their corresponding dimension (see Appendix~\ref{appd:packing}). 
This advantage has motivated the recent consideration of $D_n$ and $E_8$-related periodic boundary conditions for high-dimensional numerical simulations~\cite{berthier2020finite,biroli2020unifying}. 
For a same computational cost, these periodic boxes have a larger inscribed radius than hypercubes and thus present less pronounced finite-size corrections. 
Considering these lattices may thus help suppress obfuscating pre-asymptotic corrections to percolation criticality~\cite{mertens2018percolation,biroli2019dynamics}, especially around the upper critical dimension $n_\mathrm{u}$. Further interest in $D_n$ lattices also stems from its inclusion of the canonical three-dimensional face-centered cubic lattice, $D_3$.

In this work, we investigate the two canonical lattice percolation thresholds in $D_n$ for $n=3$ to $13$ as well as for $E_8$-related  in $n=6 \sim 9$. In Section~\ref{sec:series}, we first derive the series expansion for both $p_\mathrm{c}^\mathrm{site}$ and $p_\mathrm{c}^\mathrm{bond}$ on $D_n$ lattices based on lattice animal enumeration. We then describe the invasion percolation algorithm in Section~\ref{sec:ivd}, and analyze the numerical percolation results in Section~\ref{sec:results}. We briefly conclude in Section~\ref{sec:conclusion}.

\section{Series expansion}
\label{sec:series}

In this section we derive high-dimensional series expansions for both site and bond percolation thresholds on $D_n$ lattices by counting lattice animals embedded on the lattice~\cite{lunnon1975counting,mertens2018series}. 
For site percolation a \emph{site animal} of size $v$ is a cluster of $v$ lattice vertices connected after connecting all neighboring vertex pairs. 
Similarly, for bond percolation a \emph{bond animal} of size $e$ consists of a connected set of $e$ lattice edges.
In both cases, the perimeter $t$ is the number of incident vertices (or edges) for the lattice animal but not part of it.
Two lattice animals are distinct if they do not overlap through translation.
We here denote the number of site and bond animals of perimeter $t$ on a $n$-dimensional lattice as $g_v^t(n)$ and $g_e^t(n)$, respectively, where both $t$ and $g$ are functions (polynomials in $D_n$ lattice) in terms of $n$ with a functional mapping $t \mapsto g$.

\subsection{Site percolation}

We first consider the site percolation threshold using site animals. Following Mertens \emph{et al.}~\cite{mertens2018series} we define the polynomial
\begin{equation}
A_v(q) = \sum_t g_v^t q^t, 
\end{equation}
in terms of $q=1-p$. In particular, $A_v(1) \equiv A_v$ gives the total number of lattice animals of size $v$ in an $n$-dimensional lattice. At covering fraction $p$, the expected site cluster size on the lattice is then
\begin{equation} \label{eq:sitesize}
S = \sum_v v^2 p^{v-1} A_v(1-p) \equiv \sum_{\ell=0}^\infty b_\ell (n) p^\ell,
\end{equation}
where we have expanded $S$ as a power series in $p$. Because $A_v(q)$ has a factor of $p^{v-1}$, obtaining $b_\ell$ only requires $A_1,..., A_{\ell+1}$, i.e., counting $g_1^t$ to $g_{\ell+1}^t$. Once these terms are known, $p_\mathrm{c}^\mathrm{site}$ can be approximated by re-summing the terms using a Pad\'e approximant $b_{\ell-1}/b_{\ell}$. 

The objective is thus to count $g_v^t$ and express it as a polynomial in $n$. On hypercubic lattices the computational cost of this enumeration is greatly simplified by the introduction of \emph{proper dimension} for lattice animals~\cite{lunnon1975counting,mertens2018series}, but this approach is not obviously generalizable for  $D_n$  lattices. 
We instead implement a more generic, brute-force algorithm~\cite{mertens1990lattice}, which traverses every possible lattice animal via a breadth first search (BFS) of the lattice vertices. 

Starting at the origin, we add every nearest-neighbor site (as described in Appendix~\ref{appd:packing}) to the perimeter set. In that set, we then choose one site and add it to the site animal set according to the following criteria:
\begin{enumerate}
\item if the coordinates lexicographically greater than the origin; 
\item if the site is newly added to the perimeter set at the previous iteration, or lexicographically greater than all sites in the site animal set.
\end{enumerate} 
These two conditions guarantee that a site animal -- after properly accounting for translational invariance -- is counted exactly once.
Once a new site is selected, the perimeter set is updated with nearest neighbors of this site, with new sites being selected until the pre-assigned size $v$ is reached. 
The perimeter $t$ of each generated lattice animal is also calculated.
Therefore, by running the algorithm once with assigned $v$ and $n$, a series of integer values of $(t(n), g_v^t(n) )$ can be obtained.

The next step entails obtaining the analytical polynomial form, $t(n)$ and $g_v^t( n )$. 
We first consider $t$. Knowing that the polynomial form of $t(n)$ has the same leading order as the vertex connectivity, i.e., $t = 2n(n-1) + \OO(n)$ which is quadratic, we can relate the site animals in different dimensions by a linear fit of $t$,
\begin{equation}
t = 2n(n-1)v + c^t_1 n+ c^t_0.
\end{equation}
The coefficients $c^t_0$ and $c^t_1$ can then be extracted with $t$ results from two different dimensions. 

The polynomial $g_v^t( n )$ is also obtained by solving a linear system. The (upper bound of the) order of this polynomial must, however, be determined in advance.
Because the total number of lattice animals is $\sim [2n(n-1)]^{v-1}$, the order of $g_v^t( n )$ is also at most $n^{2(v-1)}$. And because the orientational degeneracy under $D_n$ symmetry requires that $g_v^t( n )$ always has roots $n(n-1)$, the order is further reduced to $n^{2(v-2)}$. Therefore, we require the numerical $g_v^t( n )$ results for at most $2v - 3$ different dimensions and solve the equation
\begin{equation} \label{eq:gnsite}
g_v^t( n ) / [n(n-1)] = \sum_{i=0}^{2(v-2)} c^g_i n^i,
\end{equation}
to obtain $c^g_i$. Results for $t$ and $g$ polynomials are available in Ref.~\onlinecite{lpdata}. 
While the validity of this fitting form has yet to be mathematically demonstrated, the correctness of $g_v^t$ polynomials can be empirically tested by checking that the residual vanishes when fitting the results of a (larger-than-necessary) number of dimensions. 
We have evaluated site animals up to dimension $n=15$, which is sufficient to solve Eq.~\eqref{eq:gnsite} in $v \le 6$.
However, because the total number of site animals, $A_{v} \sim n^{2(v-1)}$, grows exponentially with $v$, obtaining results for $v > 6$ lies beyond current computational reach. 

Having $t$ and $g_v^t$ with $v \le 6$, we obtain the first six terms in the expansion for $S$ (Eq.~\eqref{eq:sitesize})
\begin{equation} \begin{aligned}
b_0 &= 1, \\
b_1 &= 2 n (n - 1), \\
b_2 &= 2 n (n - 1) (2 n^2 - 6 n + 7),  \\
b_3 &= 2 n (n - 1)(4 n^4 - 24 n^3 + 57 n^2 -53 n + 12), \\
b_4 &= 2 n (n - 1)(8 n^6 - 72 n^5 + 272 n^4 - 552 n^3  + 804 n^2 \\ &\quad - 1102 n + 857), \\
b_5 &= 2 n (n - 1)(16 n^8 - 192 n^7 + 1004 n^6 - 3028 n^5 + \\ & 6018 n^4 - \frac{17710}{3} n^3 - 11851 n^2 +\frac{284075}{6} n -43202).
\end{aligned} \end{equation}
The approximant $b_0/b_1$ agrees with the threshold of the Bethe lattice, i.e., a branching tree of degree $z = 2 n (n-1)$ at leading order ~\cite{stauffer1994percolation},
\begin{equation}
\label{eq:Bethe}
p_\mathrm{c,Bethe} = \frac{1}{z - 1} \equiv \frac{1}{\sigma}.
\end{equation}
In the following we denote $1/\sigma$ the Bethe lattice limit of the percolation threshold on a lattice.

In general, $b_\ell$ has a leading order of $n^{2 \ell}$, and $b_{\ell-1}/b_\ell$ provides an approximation for $p_\mathrm{c}$ with an error that vanishes asymptotically as $\OO(n^{-(\ell+2)})$. For comparison, $b_\ell \sim n^{\ell}$ for a hypercubic lattice and the approximant $b_{2\ell }/b_{2\ell +1}$ have the same order of error $\OO(n^{-(\ell+2)})$~\cite{mertens2018series}.
Expanding $b_4/b_5$, in particular, gives
\begin{equation} \label{eq:expansionsite}
p_\mathrm{c}^\mathrm{site} = \frac{1}{\sigma} + \frac{1}{n^3} + \frac{23}{8 n^4} + \frac{17}{2 n^5} + \frac{999}{32 n^6} + \OO(n^{-7}).
\end{equation}
The accuracy of this series is evaluated in Sec.~\ref{sec:pccompare}.

\subsection{Bond percolation}

For the bond percolation, we similarly define the bond polynomial
\begin{equation}
A_e(q) = \sum_t g_e^t q^t, 
\end{equation}
which gives the expected bond cluster size
\begin{equation} \label{eq:bondsize}
S = \sum_e e^2 p^{e-1} A_e(1-p) \equiv \sum_{\ell=0}^\infty b_\ell (n) p^\ell
\end{equation}
as a polynomial in $p$. 
The enumeration scheme for bond animals is essentially the same as for site animals, with the exception that we now maintain bonds, which are indexed as the coordinates of the lexicographically smaller vertex on this bond, in addition to the orientation index -- from 1 to $n(n-1)$ -- of the bond. 
The bond animal enumeration is then used to obtain a series of numerical values $(t(n), g_e^t(n) )$. The perimeter polynomial $t(n)$ for bond animal is also quadratic with $n$, but the leading prefactor is not fixed. The polynomial is thus obtained by fitting $t(n)$ in at least three dimensions. 
A bond animal of size $e$ includes at most $e+1$ sites, hence the order of $g_e^t(n)$ is at most $n^{2 e}$, including roots $n(n-1)$.
This leads to $2 e - 1$ different dimensions being required for solving the linear equation for  $g_e^t(n)$, similar to Eq.~\eqref{eq:gnsite},
\begin{equation} \label{eq:gnbond}
g_e^t(n) / [n(n-1)] = \sum_{i=0}^{2(e-1)} c^g_i n^i.
\end{equation}
Bond animals can thus be evaluated up to dimension $n=12$ and Eq.~\eqref{eq:gnbond} can be solved up to $e=5$. Results for $t$ and $g$ polynomials are also available in Ref.~\onlinecite{lpdata}. Here as well, because the total number of bond animals $A_{e} \sim n^{2 e}$ grows exponentially with $e$, results for $e > 5$ lie beyond current computational reach. 

Invoking Eq.~\eqref{eq:bondsize} we obtain
\begin{equation} \begin{aligned}
b_0 &= n (n - 1), \\
b_1 &= 2 n (n - 1)(2 n^2 - 2 n - 1), \\
b_2 &= 2 n (n - 1) (4 n^4 - 8 n^3 + 9),  \\
b_3 &= 2 n (n - 1)(8 n^6 - 24 n^5 + 12 n^4  - 8 n^3 \\ &\quad + 27 n^2 + 131 n - 218), \\
b_4 &= 2 n (n - 1) (16 n^8 - 64 n^7 + 64 n^6 - 48 n^5 \\ &\quad + 56 n^4 + 328 n^3 + 1534 n^2 - 7778 n + 7499).
\end{aligned} \end{equation}
Note that we have $b_\ell \sim \OO(n^{2 \ell + 2})$ which is two orders (in $n$) higher than for site percolation. 
Note also that unlike for site percolation, $b_0/b_1$ here has yet to converge to the Bethe lattice limit at leading order.
For $\ell \ge 2$, however, the Pad\'e approximant $p_\mathrm{c}^\mathrm{bond} \approx b_{\ell-1}/b_\ell$ has an error of $\OO(n^{-(\ell+3)})$ one order smaller than for $p_\mathrm{c}^\mathrm{site}$. 
In particular, expanding $b_3/b_4$ gives
\begin{equation} \label{eq:expansionbond}
p_\mathrm{c}^\mathrm{bond} = \frac{1}{\sigma} + \frac{1}{n^5} + \frac{81}{16 n^6} + \OO(n^{-7}).
\end{equation}
The accuracy of this series is also evaluated  in Sec.~\ref{sec:pccompare}.

\section{Invasion percolation}
\label{sec:ivd}
In this section we briefly describe the invasion percolation algorithm by Mertens and Moore~\cite{mertens2018percolation} (derived from Ref.~\onlinecite{wilkinson1983invasion}) for an arbitrary lattice structure, and then analyze its complexity for the considered lattices.

As stated in the introduction, the algorithm grows a single cluster without explicitly storing the lattice grid.
Two data structures are then used: a set (collection of unique elements) $\mathcal{S}$ to maintain all sites (or bonds) belonging to the cluster as well as those incident to them; and a priority queue $\mathcal{Q}$  for the stepwise growth of the cluster.
For site percolation, starting from the origin every neighboring vertex is inserted (following Appendix~\ref{appd:packing}) into $\mathcal{S}$.
For each of these new vertices, a random weight $w_i \in [0, 1)$ is assigned and the vertex is inserted into $\mathcal{Q}$ with $w_i$ as the key. 
For the next step, the vertex of minimum weight in $\mathcal{Q}$ is popped, incrementing the cluster size $N$. The previous steps are repeated until the pre-assigned cluster size $N=N_0$ is attained. The expected set size at a certain $N$, denoted $B(N) = \langle |\mathcal{S}(N) | \rangle$, is computed by averaging the set size among independent realizations.
For the bond percolation, we start with an arbitrary bond incident to the origin and otherwise the same procedure is used.

The cluster obtained by invasion percolation process simultaneously approaches the giant component at $p_\mathrm{c}$ with the scaling form~\cite{mertens2017percolation,mertens2018percolation}
\begin{equation} \label{eq:pcinvade}
\frac{N}{B(N)} \approx p_\mathrm{c} ( 1 - c N^{-\delta} )
\end{equation}
where $\delta$ is the correction exponent and $c$ a fitting constant. 

For each instance, the space complexity is
\[
n|\mathcal{S}| + |\mathcal{Q}| \sim n N/p_\mathrm{c} \sim O(n \sigma N)
\]
where the factor of $n$ accounts for the size of an $n$-dimensional vector. For $D_n$ lattice the space complexity is thus $\OO(n^3 N)$.
Although the space complexity is larger, by a factor of $n$, than for $\mathbb{Z}^n$ lattices ($\OO(n^2 N)$), the memory requirement is still moderate for contemporary computers.
The time complexity depends on the complexity of the insertion to $\mathcal{Q}$ which is at most $\OO(n + \log \mathcal{|Q|}) \approx \OO(\log N)$, and thus  $\OO(N \log N)$ in total.
In practice, we can grow clusters up to $N_0 = 1.5 \times 10^7$ in $n=3$ and up to $2 \times 10^5$ in $n=13$, within a memory usage of less than 10 GB.
At least $10^4$ independent clusters are obtained for each lattice, which results in each realization is usually taking less than a minute on an AMD Ryzen 3900x processor. 

\section{Result and discussion}
\label{sec:results}

In this section we compare the numerical threshold values obtained from the invasion percolation described in Sec.~\ref{sec:ivd} with the series expansion results obtained in Sec.~\ref{sec:series}.

\subsection{Numerical thresholds}
\label{sec:pcnumerical}

\begin{figure}
\includegraphics[width=0.98\columnwidth]{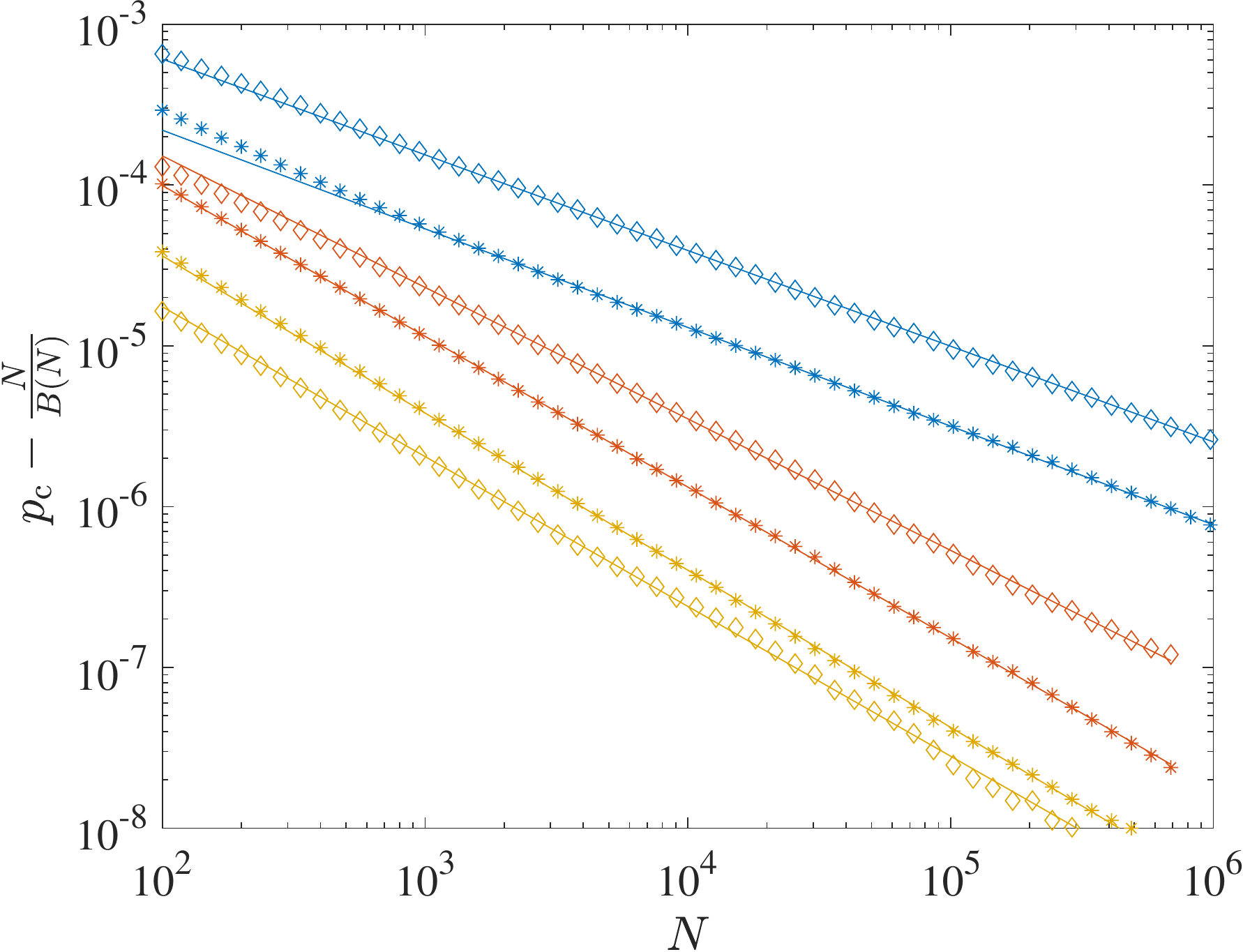}
\caption{Convergence of $N/B(N)$ to site (diamonds) and bond (asterisks) percolation thresholds on $D_n$ lattices in $n=4$ (blue), $8$ (red) and $12$ (yellow). Finite-size correction scales as $\OO(N^{-\delta})$ (solid lines), where $\delta \rightarrow 1$ as $n\rightarrow\infty$.}
\label{fig:conv}
\end{figure}

Table~\ref{tab:pc} reports both site and bond percolation thresholds for $D_n$ as well as for $E_8$-related lattices obtained by fitting the numerical $N/B(N)$ results with Eq.~\ref{eq:pcinvade}.
The values for $n=3$-$6$ are consistent with published values, and, except for $n=3$, our results are at least an order of magnitude more accurate. For $ 6<n \le 13$ no prior result is known. As in Ref.~\onlinecite{mertens2018percolation} for $\mathbb{Z}^n$ lattices, $p_\mathrm{c}$ results for $D_n$ lattices are obtained with higher precision -- for comparable computational efforts -- as $n$ increases (Fig.~\ref{fig:conv}).
Because the correction exponent $\delta$ (Eq.~\eqref{eq:pcinvade}) increases with $n$, finite-size corrections then decay faster. For the range of $n$ considered, this advantage compensates for the decrease in $N_0$ imposed by the growing memory cost. As a result, a relative uncertainty of $10^{-6}$ to $10^{-7}$ is obtained for all investigated dimensions.

\begin{table}
\caption{Site and bond percolation threshold on $D_n$ and $E_8$ related lattices}
\label{tab:pc}
\begin{tabular}{cll}
\hline
Lattice & $p_\mathrm{c}^\mathrm{site}$ & $p_\mathrm{c}^\mathrm{bond}$ \\
\hline
$D_3$ & 0.199 236(4) &  0.120 162 0(8) \\
& 0.199 235 17(20)~\cite{xu2014simultaneous} & 0.120 163 5(10)~\cite{lorenz1998precise} \\
$D_4$ & 0.084 200 1(11) & 0.049 519 3(8)  \\
& 0.084 10(23)~\cite{kotwica2019efficient} & 0.049 517(1)~\cite{xun2020precise} \\
$D_5$ & 0.043 591 3(6) & 0.027 181 3(2)  \\
& 0.043 1(3)~\cite{van1998calculation} & 0.026(2)~\cite{van1998calculation} \\
$D_6$ & 0.026 026 74(12) &  0.017 415 56(5) \\
& 0.025 2(5)~\cite{van1998calculation} & \\
$D_7$ & 0.017 167 30(5)& 0.012 217 868(13)  \\
$D_8$ & 0.012 153 92(4) & 0.009 081 804(6) \\
$D_9$ & 0.009 058 70(2) & 0.007 028 457(3) \\
$D_{10}$ & 0.007 016 353(9) & 0.005 605 579(6) \\
$D_{11}$ & 0.005 597 592(4) & 0.004 577 155(3) \\
$D_{12}$ & 0.004 571 339(4) & 0.003 808 960(2) \\
$D_{13}$ & 0.003 804 565(3) & 0.003 219 701 3(14) \\
\hline
$E_6$ & 0.021 940 21(14) & 0.014 432 05(8) \\
$E_7$ & 0.011 623 06(4) & 0.008 083 68(2) \\
$E_8$ & 0.005 769 91(2) & 0.004 202 07(2) \\
$\Lambda_9$ & 0.004 808 39(2) & 0.003 700 865(11) \\
\hline
\end{tabular}
\end{table}

\begin{figure}
\includegraphics[width=0.98\columnwidth]{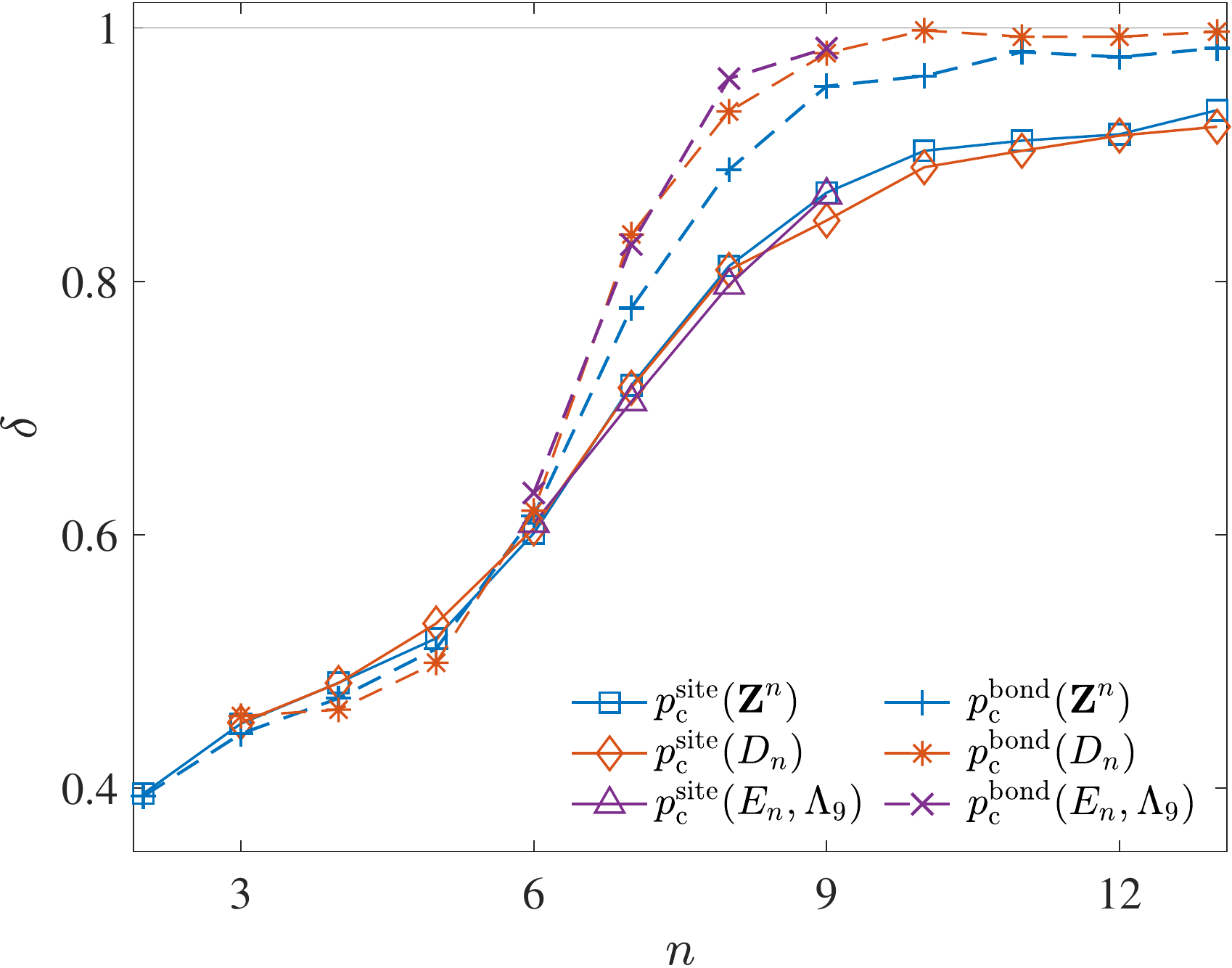}
\caption{Finite-size scaling exponent $\delta$ for site and bond percolation on $\mathbb{Z}^n$, $D_n$ and $E_8$-related lattices, $E_6, E_7, E_8$ and $\Lambda_9$. (Results for $\mathbb{Z}^n$ site percolation in $n=4$ to $13$ are from Ref.~\onlinecite{mertens2018percolation}.) Error bars from fitting are smaller than (comparable to) the marker size in $n \le 6$ (in $n > 6$). Lines are guides to the eye. Note that $\delta$ generically grows with $n$ but its value is not universal.}
\label{fig:delta}
\end{figure}

Because $\delta$ controls the convergence rate of invasion percolation, it is interesting to compare its behavior for different lattices. 
As a first glance, $\delta$ increases with $n$ for both $\mathbb{Z}^n$ and $D_n$ lattices and tends to $1$ as dimension increases, as expected from the Bethe lattice analysis~\cite{mertens2017percolation,mertens2018percolation}. 
While for site percolation on $\mathbb{Z}^n$, $D_n$ and $E_8$-related lattices $\delta$ appears similar, the exponent evolves differently for bond percolation on different lattices as well as for either type of percolation on a same lattice. 
Because the exact value of $\delta$ depends on the type of percolation as well as on lattice geometry, we conclude that the exponent is not universal.  
As a corollary, $\delta$ may be a useful quantity for selecting a lattice for studying criticality; a greater $\delta$ indeed implies a faster decay of certain finite-size corrections.

\subsection{Comparison with series expansion}
\label{sec:pccompare}

\begin{figure}
\includegraphics[width=0.98\columnwidth]{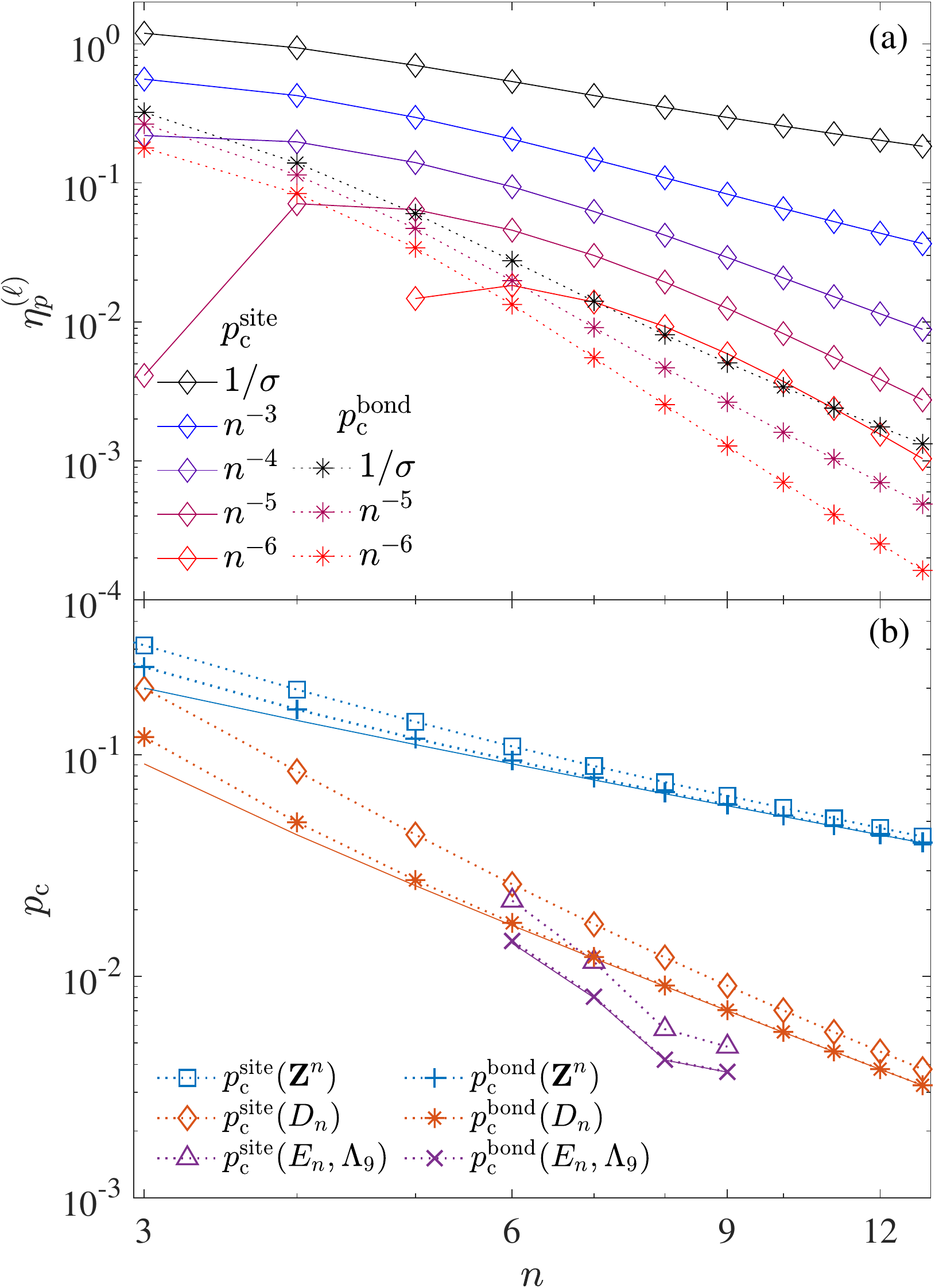}
\caption{(a) Relative error for the site (diamonds) and bond (asterisks) percolation thresholds on $D_n$ lattices predicted by series expansion for various highest-order terms. Note that for site percolation the high-order lines are truncated in small $n$ because the relative error then changes sign. Lines are guides to the eye. (b) Percolation thresholds on $\mathbb{Z}^n$, $D_n$ and $E_8$-related lattices (markers with dotted line) compared to the Bethe lattice limit $1/\sigma$ (solid line), which matches the bond percolation threshold well in all three lattice types.}
\label{fig:pcseries}
\end{figure}

Our precise numerical thresholds for $D_n$ lattices can be compared with the series prediction obtained for both site and bond percolation in Sec.~\ref{sec:series}. The relative error of the expansion up to $n^{-\ell}$ term, defined as
\begin{equation}
\eta_p^{(\ell)} = \left| p_\mathrm{c,simulation} - p_\mathrm{c,series}^{(\ell)} \right| / p_\mathrm{c,simulation} ,
\end{equation}
is shown in Fig.~\ref{fig:pcseries}(a). 
As expected, these thresholds converges gradually to the Bethe lattice value, $1/\sigma$, in the large $n$ limit. For site percolation, this convergence rate is fairly slow -- a $\sim 10\%$ deviation persists even in $n=13$ -- but introducing higher-order terms in the series dramatically reduces that error. Including terms of order up to $n^{-6}$ leads to a relative error of $\sim 0.1\%$ in $n=13$. 
For bond percolation, because the prefactors for both $n^{-3}$ and $n^{-4}$ in the expansion form are zero, the deviation is already down to $\sim 0.1 \%$ in $n=13$. Including two more terms in Eq.~\eqref{eq:expansionbond} further divides the error by a factor $\sim n^{2}$. 
The series expansion in Eqs.~\eqref{eq:expansionsite} and~\eqref{eq:expansionbond} is thus expected to predict percolation thresholds with very high accuracy for $n > 13$.

Percolation thresholds for $\mathbb{Z}^n$, $D_n$ and $E_8$-related lattices are compared with the Bethe lattice result in Eq.~\eqref{eq:Bethe}.
(Although a dimensional series expansion is not available for $E_8$-related lattices, the site connectivity, $z=72, 126, 240, 272$ for $E_6, E_7, E_8$ and $\Lambda_9$ lattices~\cite{Conway1988}, respectively, alone suffices for this comparison (see Appendix~\ref{appd:packing}).) 
In all three cases, the Bethe lattice prediction better matches the bond than the site percolation threshold (Fig.~\ref{fig:pcseries}(b)).
For $\mathbb{Z}^n$ and $D_n$ lattices this result is expected from the series expansion. In the large $n$ limit, the deviation of $p_\mathrm{c}^\mathrm{bond}$ from the Bethe lattice limit is of $\OO(n^{-2}), \OO(n^{-3}), \OO(n^{-3})$ and $ \OO(n^{-5})$ for $p_\mathrm{c}^\mathrm{site}(\mathbb{Z}^n), p_\mathrm{c}^\mathrm{bond}(\mathbb{Z}^n), p_\mathrm{c}^\mathrm{site}(D_n)$ and $p_\mathrm{c}^\mathrm{bond}(D_n)$, respectively.
For $E_8$-related lattices, for which no such series exist, the same trend is observed. More specifically, the deviation is $\lesssim 1\%$ for bond percolation and $\lesssim 40\%$ for site percolation. This concordance suggests that the effect might be more than a mere coincidence. Yet it lacks a physical explanation. A generic scaling form for the percolation threshold beyond the Bethe lattice approximation might be informative in this respect, but is still found lacking.

\section{Conclusion}
\label{sec:conclusion}

We have reported the series expansion and numerical percolation thresholds for $D_n$ lattices as well as the numerical thresholds for $E_8$-related lattices from $n=6$ to $9$. 
The excellent agreement between the two independent approaches cross-validates their results. 
Remarkably, bond percolation presents much faster decaying finite-size corrections than site percolation for invasion percolation in $n > 6$. This finding suggests that pre-asymptotic corrections might be most efficiently suppressed in the former.
The Bethe lattice approximation to the percolation threshold also presents a markedly higher precision for bond percolation than for site percolation for $D_n$, due to the vanishing of the first subleading order coefficients in the series expansion.  
This feature thus appears to be generic for lattices other than $\mathbb{Z}^n$, for which it was first reported~\cite{gaunt1976percolation,gaunt1978bond,mertens2018series}.
Our finding identify unresolved features of percolation and set the stage for investigating percolation criticality on high-dimensional lattices beyond the conventional hypercubic geometry.

\begin{acknowledgements}
We thank R.~M.~Ziff for carefully maintaining the Percolation threshold Wikipedia page, which has greatly facilitated our literature search. This work was supported by a grant from the Simons Foundation (\#454937). The computations were carried out on the Duke Compute Cluster and Open Science Grid~\cite{osg07,osg09}, supported by National Science Foundation award 1148698, and the U.S. Department of Energy's Office of Science. Data relevant to this work have been archived and can be accessed at the Duke Digital Repository~\cite{lpdata}.
\end{acknowledgements}

\appendix

\section{Lattice packing}
\label{appd:packing}

In this appendix we briefly review the structure of the high-dimensional lattices considered in this study, following the construction in Ref.~\onlinecite{convay1982fast}. As reference, the conventional $n$-dimensional hypercubic lattices, $\mathbb{Z}^n$, is defined as a set of $n$-dimensional vectors of integer components. The nearest-neighbor vector $(a_1, a_2, ..., a_n)$ in $\mathbb{Z}^n$ are $(\pm 1, 0^{n-1})$ (this notation means $(\pm 1, 0, ...)$ and their permutations). The number of nearest neighbors (kissing number) is thus $2n$. 
$D_n$ lattices can be viewed as a subset of $\mathbb{Z}^n$ in which the coordinates have even sum. The nearest neighbor vectors are $(\pm1^2, 0^{n-2})$, thus resulting in $2n(n-1)$ nearest neighbors in total. In $n=3$, for example, the 12 nearest-neighbor vectors for the $D_3 \equiv \mathrm{fcc}$ lattice read
\begin{equation*}
\begin{matrix}
(1, 1, 0), & (1, -1, 0), & (-1, 1, 0), & (-1, -1, 0), \\
(1, 0, 1), & (1, 0, -1), & (-1, 0, 1), & (-1, 0, -1), \\
(0, 1, 1), & (0, 1, -1), & (0, -1, 1), & (0, -1, -1). \\
\end{matrix}
\end{equation*}

$D_3$, $D_4$ and $D_5$ lattices are the densest packings of equal spheres in the corresponding dimensions. 
The densest sphere packings for $n=6$ to $9$ are $E_6$, $E_7$, $E_8$ and $\Lambda_9$ lattices, respectively.
In particular, the $E_8$ lattice consists of two $D_8$ lattice points with offset $(\frac{1}{2}^8)$. The nearest-neighbor vectors of $E_8$ can be viewed as four groups,
\begin{equation} \label{eq:E8vec}
\begin{cases}
\pm (\frac{1}{2}^8), & \hfill \hfill 2~\mathrm{vectors}, \\
(\frac{1}{2}^4, -\frac{1}{2}^4), & \hfill 70~\mathrm{vectors}, \\
(\frac{1}{2}^2, -\frac{1}{2}^6)~\mathrm{and}~(-\frac{1}{2}^2, \frac{1}{2}^6), &\hfill 56~\mathrm{vectors}, \\
(\pm 1^2, 0^6), & 112~\mathrm{vectors},
\end{cases}
\end{equation}
and thus each vertex has $240$ nearest neighbors in total.
$E_7$ lattice is a cross-section of $E_8$ in $n=7$. One of the choices to generate nearest neighbor vectors in $E_7$ is the subset of Eq.~\eqref{eq:E8vec} with zero sum, which results in $126$ vectors in total. 
Further constraining $a_1 + a_8 = 0$ leads to $72$ nearest neighbor vectors in $E_6$.
The $\Lambda_9$ lattice is not unique, but one of its forms can be constructed similarly to $E_8$. It consists of two $D_9$ lattice points, offset by $(\frac{1}{2}, ..., \frac{1}{2}, 0)$, which results in $272$ nearest-neighbor vectors. 
Note that in $d=10$ the (presumed) densest packing is a non-lattice~\cite{conway1995all}, and thus $n=9$ offers a natural end to our consideration of dense packing lattices.

Finally, we note that in some dimensions there exist structures comparable to lattices considered here. For example, in $n=3$ face-centered cubic ($D_3$) is strongly related to the hexagonal closed-packed structure. 
As a result, their site percolation thresholds are close although not identical~\cite{lorenz2000similarity}. In $n=5, 6, 7$, four similar structures are known~\cite{conway1995all} and their $p_\mathrm{c}$ values may also differ marginally. In $n=9$ a continuum of structures can be constructed. In the current study these alternative lattices were not considered, hence their constructions are omitted from this appendix.

\bibliography{abbrev}

\begin{thebibliography}{32}%
\makeatletter
\providecommand \@ifxundefined [1]{%
 \@ifx{#1\undefined}
}%
\providecommand \@ifnum [1]{%
 \ifnum #1\expandafter \@firstoftwo
 \else \expandafter \@secondoftwo
 \fi
}%
\providecommand \@ifx [1]{%
 \ifx #1\expandafter \@firstoftwo
 \else \expandafter \@secondoftwo
 \fi
}%
\providecommand \natexlab [1]{#1}%
\providecommand \enquote  [1]{``#1''}%
\providecommand \bibnamefont  [1]{#1}%
\providecommand \bibfnamefont [1]{#1}%
\providecommand \citenamefont [1]{#1}%
\providecommand \href@noop [0]{\@secondoftwo}%
\providecommand \href [0]{\begingroup \@sanitize@url \@href}%
\providecommand \@href[1]{\@@startlink{#1}\@@href}%
\providecommand \@@href[1]{\endgroup#1\@@endlink}%
\providecommand \@sanitize@url [0]{\catcode `\\12\catcode `\$12\catcode
  `\&12\catcode `\#12\catcode `\^12\catcode `\_12\catcode `\%12\relax}%
\providecommand \@@startlink[1]{}%
\providecommand \@@endlink[0]{}%
\providecommand \url  [0]{\begingroup\@sanitize@url \@url }%
\providecommand \@url [1]{\endgroup\@href {#1}{\urlprefix }}%
\providecommand \urlprefix  [0]{URL }%
\providecommand \Eprint [0]{\href }%
\providecommand \doibase [0]{http://dx.doi.org/}%
\providecommand \selectlanguage [0]{\@gobble}%
\providecommand \bibinfo  [0]{\@secondoftwo}%
\providecommand \bibfield  [0]{\@secondoftwo}%
\providecommand \translation [1]{[#1]}%
\providecommand \BibitemOpen [0]{}%
\providecommand \bibitemStop [0]{}%
\providecommand \bibitemNoStop [0]{.\EOS\space}%
\providecommand \EOS [0]{\spacefactor3000\relax}%
\providecommand \BibitemShut  [1]{\csname bibitem#1\endcsname}%
\let\auto@bib@innerbib\@empty
\bibitem [{\citenamefont {Stauffer}\ and\ \citenamefont
  {Aharony}(1994)}]{stauffer1994percolation}%
  \BibitemOpen
  \bibfield  {author} {\bibinfo {author} {\bibfnamefont {D.}~\bibnamefont
  {Stauffer}}\ and\ \bibinfo {author} {\bibfnamefont {A.}~\bibnamefont
  {Aharony}},\ }\href@noop {} {\emph {\bibinfo {title} {Introduction To
  Percolation Theory}}}\ (\bibinfo  {publisher} {Taylor \& Francis},\ \bibinfo
  {year} {1994})\BibitemShut {NoStop}%
\bibitem [{\citenamefont {Mertens}\ and\ \citenamefont
  {Moore}(2018{\natexlab{a}})}]{mertens2018percolation}%
  \BibitemOpen
  \bibfield  {author} {\bibinfo {author} {\bibfnamefont {S.}~\bibnamefont
  {Mertens}}\ and\ \bibinfo {author} {\bibfnamefont {C.}~\bibnamefont
  {Moore}},\ }\href {\doibase 10.1103/PhysRevE.98.022120} {\bibfield  {journal}
  {\bibinfo  {journal} {Phys. Rev. E}\ }\textbf {\bibinfo {volume} {98}},\
  \bibinfo {pages} {022120} (\bibinfo {year} {2018}{\natexlab{a}})}\BibitemShut
  {NoStop}%
\bibitem [{\citenamefont {Huang}\ \emph {et~al.}(2018)\citenamefont {Huang},
  \citenamefont {Hou}, \citenamefont {Wang}, \citenamefont {Ziff},\ and\
  \citenamefont {Deng}}]{huang2018critical}%
  \BibitemOpen
  \bibfield  {author} {\bibinfo {author} {\bibfnamefont {W.}~\bibnamefont
  {Huang}}, \bibinfo {author} {\bibfnamefont {P.}~\bibnamefont {Hou}}, \bibinfo
  {author} {\bibfnamefont {J.}~\bibnamefont {Wang}}, \bibinfo {author}
  {\bibfnamefont {R.~M.}\ \bibnamefont {Ziff}}, \ and\ \bibinfo {author}
  {\bibfnamefont {Y.}~\bibnamefont {Deng}},\ }\href {\doibase
  10.1103/PhysRevE.97.022107} {\bibfield  {journal} {\bibinfo  {journal} {Phys.
  Rev. E}\ }\textbf {\bibinfo {volume} {97}},\ \bibinfo {pages} {022107}
  (\bibinfo {year} {2018})}\BibitemShut {NoStop}%
\bibitem [{\citenamefont {Biroli}\ \emph {et~al.}(2019)\citenamefont {Biroli},
  \citenamefont {Charbonneau},\ and\ \citenamefont {Hu}}]{biroli2019dynamics}%
  \BibitemOpen
  \bibfield  {author} {\bibinfo {author} {\bibfnamefont {G.}~\bibnamefont
  {Biroli}}, \bibinfo {author} {\bibfnamefont {P.}~\bibnamefont {Charbonneau}},
  \ and\ \bibinfo {author} {\bibfnamefont {Y.}~\bibnamefont {Hu}},\ }\href
  {\doibase 10.1103/PhysRevE.99.022118} {\bibfield  {journal} {\bibinfo
  {journal} {Phys. Rev. E}\ }\textbf {\bibinfo {volume} {99}},\ \bibinfo
  {pages} {022118} (\bibinfo {year} {2019})}\BibitemShut {NoStop}%
\bibitem [{\citenamefont {Xun}\ and\ \citenamefont
  {Ziff}(2020{\natexlab{a}})}]{xun2020precise}%
  \BibitemOpen
  \bibfield  {author} {\bibinfo {author} {\bibfnamefont {Z.}~\bibnamefont
  {Xun}}\ and\ \bibinfo {author} {\bibfnamefont {R.~M.}\ \bibnamefont {Ziff}},\
  }\href {\doibase 10.1103/PhysRevResearch.2.013067} {\bibfield  {journal}
  {\bibinfo  {journal} {Phys. Rev. Research}\ }\textbf {\bibinfo {volume}
  {2}},\ \bibinfo {pages} {013067} (\bibinfo {year}
  {2020}{\natexlab{a}})}\BibitemShut {NoStop}%
\bibitem [{\citenamefont {Wierman}(2002)}]{wierman2002accuracy}%
  \BibitemOpen
  \bibfield  {author} {\bibinfo {author} {\bibfnamefont {J.~C.}\ \bibnamefont
  {Wierman}},\ }\href {\doibase 10.1103/PhysRevE.66.027105} {\bibfield
  {journal} {\bibinfo  {journal} {Phys. Rev. E}\ }\textbf {\bibinfo {volume}
  {66}},\ \bibinfo {pages} {027105} (\bibinfo {year} {2002})}\BibitemShut
  {NoStop}%
\bibitem [{\citenamefont {Lorenz}\ and\ \citenamefont
  {Ziff}(1998)}]{lorenz1998precise}%
  \BibitemOpen
  \bibfield  {author} {\bibinfo {author} {\bibfnamefont {C.~D.}\ \bibnamefont
  {Lorenz}}\ and\ \bibinfo {author} {\bibfnamefont {R.~M.}\ \bibnamefont
  {Ziff}},\ }\href {\doibase 10.1103/PhysRevE.57.230} {\bibfield  {journal}
  {\bibinfo  {journal} {Phys. Rev. E}\ }\textbf {\bibinfo {volume} {57}},\
  \bibinfo {pages} {230} (\bibinfo {year} {1998})}\BibitemShut {NoStop}%
\bibitem [{\citenamefont {Xu}\ \emph {et~al.}(2014)\citenamefont {Xu},
  \citenamefont {Wang}, \citenamefont {Lv},\ and\ \citenamefont
  {Deng}}]{xu2014simultaneous}%
  \BibitemOpen
  \bibfield  {author} {\bibinfo {author} {\bibfnamefont {X.}~\bibnamefont
  {Xu}}, \bibinfo {author} {\bibfnamefont {J.}~\bibnamefont {Wang}}, \bibinfo
  {author} {\bibfnamefont {J.-P.}\ \bibnamefont {Lv}}, \ and\ \bibinfo {author}
  {\bibfnamefont {Y.}~\bibnamefont {Deng}},\ }\href {\doibase
  10.1007/s11467-013-0403-z} {\bibfield  {journal} {\bibinfo  {journal} {Front.
  Phys.}\ }\textbf {\bibinfo {volume} {9}},\ \bibinfo {pages} {113} (\bibinfo
  {year} {2014})}\BibitemShut {NoStop}%
\bibitem [{\citenamefont {Kotwica}\ \emph {et~al.}(2019)\citenamefont
  {Kotwica}, \citenamefont {Gronek},\ and\ \citenamefont
  {Malarz}}]{kotwica2019efficient}%
  \BibitemOpen
  \bibfield  {author} {\bibinfo {author} {\bibfnamefont {M.}~\bibnamefont
  {Kotwica}}, \bibinfo {author} {\bibfnamefont {P.}~\bibnamefont {Gronek}}, \
  and\ \bibinfo {author} {\bibfnamefont {K.}~\bibnamefont {Malarz}},\ }\href
  {\doibase 10.1142/S0129183119500554} {\bibfield  {journal} {\bibinfo
  {journal} {Int. J. Mod. Phys. C}\ }\textbf {\bibinfo {volume} {30}},\
  \bibinfo {pages} {1950055} (\bibinfo {year} {2019})}\BibitemShut {NoStop}%
\bibitem [{\citenamefont {Xun}\ and\ \citenamefont
  {Ziff}(2020{\natexlab{b}})}]{xun2020bond}%
  \BibitemOpen
  \bibfield  {author} {\bibinfo {author} {\bibfnamefont {Z.}~\bibnamefont
  {Xun}}\ and\ \bibinfo {author} {\bibfnamefont {R.~M.}\ \bibnamefont {Ziff}},\
  }\href {\doibase 10.1103/PhysRevE.102.012102} {\bibfield  {journal} {\bibinfo
   {journal} {Phys. Rev. E}\ }\textbf {\bibinfo {volume} {102}},\ \bibinfo
  {pages} {012102} (\bibinfo {year} {2020}{\natexlab{b}})}\BibitemShut
  {NoStop}%
\bibitem [{\citenamefont {Scullard}\ and\ \citenamefont
  {Ziff}(2008)}]{scullard2008critical}%
  \BibitemOpen
  \bibfield  {author} {\bibinfo {author} {\bibfnamefont {C.~R.}\ \bibnamefont
  {Scullard}}\ and\ \bibinfo {author} {\bibfnamefont {R.~M.}\ \bibnamefont
  {Ziff}},\ }\href {\doibase 10.1103/PhysRevLett.100.185701} {\bibfield
  {journal} {\bibinfo  {journal} {Phys. Rev. Lett.}\ }\textbf {\bibinfo
  {volume} {100}},\ \bibinfo {pages} {185701} (\bibinfo {year}
  {2008})}\BibitemShut {NoStop}%
\bibitem [{\citenamefont {Scullard}\ and\ \citenamefont
  {Ziff}(2010)}]{scullard2010critical}%
  \BibitemOpen
  \bibfield  {author} {\bibinfo {author} {\bibfnamefont {C.~R.}\ \bibnamefont
  {Scullard}}\ and\ \bibinfo {author} {\bibfnamefont {R.~M.}\ \bibnamefont
  {Ziff}},\ }\href {\doibase 10.1088/1742-5468/2010/03/P03021} {\bibfield
  {journal} {\bibinfo  {journal} {J. Stat. Mech. Theory Exp.}\ }\textbf
  {\bibinfo {volume} {2010}},\ \bibinfo {pages} {P03021} (\bibinfo {year}
  {2010})}\BibitemShut {NoStop}%
\bibitem [{\citenamefont {Jacobsen}(2015)}]{jacobsen2015critical}%
  \BibitemOpen
  \bibfield  {author} {\bibinfo {author} {\bibfnamefont {J.~L.}\ \bibnamefont
  {Jacobsen}},\ }\href {\doibase doi.org/10.1088/1751-8113/48/45/454003}
  {\bibfield  {journal} {\bibinfo  {journal} {J. Phys. A}\ }\textbf {\bibinfo
  {volume} {48}},\ \bibinfo {pages} {454003} (\bibinfo {year}
  {2015})}\BibitemShut {NoStop}%
\bibitem [{\citenamefont {Scullard}\ and\ \citenamefont
  {Jacobsen}(2020)}]{scullard2020bond}%
  \BibitemOpen
  \bibfield  {author} {\bibinfo {author} {\bibfnamefont {C.~R.}\ \bibnamefont
  {Scullard}}\ and\ \bibinfo {author} {\bibfnamefont {J.~L.}\ \bibnamefont
  {Jacobsen}},\ }\href {\doibase 10.1103/PhysRevResearch.2.012050} {\bibfield
  {journal} {\bibinfo  {journal} {Phys. Rev. Research}\ }\textbf {\bibinfo
  {volume} {2}},\ \bibinfo {pages} {012050} (\bibinfo {year}
  {2020})}\BibitemShut {NoStop}%
\bibitem [{\citenamefont {Kirkpatrick}(1976)}]{kirkpatrick1976percolation}%
  \BibitemOpen
  \bibfield  {author} {\bibinfo {author} {\bibfnamefont {S.}~\bibnamefont
  {Kirkpatrick}},\ }\href {\doibase 10.1103/PhysRevLett.36.69} {\bibfield
  {journal} {\bibinfo  {journal} {Phys. Rev. Lett.}\ }\textbf {\bibinfo
  {volume} {36}},\ \bibinfo {pages} {69} (\bibinfo {year} {1976})}\BibitemShut
  {NoStop}%
\bibitem [{\citenamefont {Mertens}\ and\ \citenamefont
  {Moore}(2017)}]{mertens2017percolation}%
  \BibitemOpen
  \bibfield  {author} {\bibinfo {author} {\bibfnamefont {S.}~\bibnamefont
  {Mertens}}\ and\ \bibinfo {author} {\bibfnamefont {C.}~\bibnamefont
  {Moore}},\ }\href {\doibase 10.1103/PhysRevE.96.042116} {\bibfield  {journal}
  {\bibinfo  {journal} {Phys. Rev. E}\ }\textbf {\bibinfo {volume} {96}},\
  \bibinfo {pages} {042116} (\bibinfo {year} {2017})}\BibitemShut {NoStop}%
\bibitem [{\citenamefont {Mertens}\ and\ \citenamefont
  {Moore}(2018{\natexlab{b}})}]{mertens2018series}%
  \BibitemOpen
  \bibfield  {author} {\bibinfo {author} {\bibfnamefont {S.}~\bibnamefont
  {Mertens}}\ and\ \bibinfo {author} {\bibfnamefont {C.}~\bibnamefont
  {Moore}},\ }\href {\doibase 10.1088/1751-8121/aae65c} {\bibfield  {journal}
  {\bibinfo  {journal} {J. Phys. A}\ }\textbf {\bibinfo {volume} {51}},\
  \bibinfo {pages} {475001} (\bibinfo {year} {2018}{\natexlab{b}})}\BibitemShut
  {NoStop}%
\bibitem [{\citenamefont {Conway}\ and\ \citenamefont
  {Sloane}(1988)}]{Conway1988}%
  \BibitemOpen
  \bibfield  {author} {\bibinfo {author} {\bibfnamefont {J.~H.}\ \bibnamefont
  {Conway}}\ and\ \bibinfo {author} {\bibfnamefont {N.~J.~A.}\ \bibnamefont
  {Sloane}},\ }\enquote {\bibinfo {title} {Certain important lattices and their
  properties},}\ in\ \href {\doibase 10.1007/978-1-4757-2016-7_4} {\emph
  {\bibinfo {booktitle} {Sphere Packings, Lattices and Groups}}}\ (\bibinfo
  {publisher} {Springer New York},\ \bibinfo {address} {New York, NY},\
  \bibinfo {year} {1988})\ pp.\ \bibinfo {pages} {94--135}\BibitemShut
  {NoStop}%
\bibitem [{\citenamefont {Berthier}\ \emph {et~al.}(2020)\citenamefont
  {Berthier}, \citenamefont {Charbonneau},\ and\ \citenamefont
  {Kundu}}]{berthier2020finite}%
  \BibitemOpen
  \bibfield  {author} {\bibinfo {author} {\bibfnamefont {L.}~\bibnamefont
  {Berthier}}, \bibinfo {author} {\bibfnamefont {P.}~\bibnamefont
  {Charbonneau}}, \ and\ \bibinfo {author} {\bibfnamefont {J.}~\bibnamefont
  {Kundu}},\ }\href {\doibase 10.1103/PhysRevLett.125.108001} {\bibfield
  {journal} {\bibinfo  {journal} {Phys. Rev. Lett.}\ }\textbf {\bibinfo
  {volume} {125}},\ \bibinfo {pages} {108001} (\bibinfo {year}
  {2020})}\BibitemShut {NoStop}%
\bibitem [{\citenamefont {Biroli}\ \emph {et~al.}(2020)\citenamefont {Biroli},
  \citenamefont {Charbonneau}, \citenamefont {Corwin}, \citenamefont {Hu},
  \citenamefont {Ikeda}, \citenamefont {Szamel},\ and\ \citenamefont
  {Zamponi}}]{biroli2020unifying}%
  \BibitemOpen
  \bibfield  {author} {\bibinfo {author} {\bibfnamefont {G.}~\bibnamefont
  {Biroli}}, \bibinfo {author} {\bibfnamefont {P.}~\bibnamefont {Charbonneau}},
  \bibinfo {author} {\bibfnamefont {E.~I.}\ \bibnamefont {Corwin}}, \bibinfo
  {author} {\bibfnamefont {Y.}~\bibnamefont {Hu}}, \bibinfo {author}
  {\bibfnamefont {H.}~\bibnamefont {Ikeda}}, \bibinfo {author} {\bibfnamefont
  {G.}~\bibnamefont {Szamel}}, \ and\ \bibinfo {author} {\bibfnamefont
  {F.}~\bibnamefont {Zamponi}},\ }\href@noop {} {\bibfield  {journal} {\bibinfo
   {journal} {arXiv preprint}\ } (\bibinfo {year} {2020})},\ \Eprint
  {http://arxiv.org/abs/2003.11179} {2003.11179} \BibitemShut {NoStop}%
\bibitem [{\citenamefont {Lunnon}(1975)}]{lunnon1975counting}%
  \BibitemOpen
  \bibfield  {author} {\bibinfo {author} {\bibfnamefont {W.~F.}\ \bibnamefont
  {Lunnon}},\ }\href {\doibase 10.1093/comjnl/18.4.366} {\bibfield  {journal}
  {\bibinfo  {journal} {Comput. J}\ }\textbf {\bibinfo {volume} {18}},\
  \bibinfo {pages} {366} (\bibinfo {year} {1975})}\BibitemShut {NoStop}%
\bibitem [{\citenamefont {Mertens}(1990)}]{mertens1990lattice}%
  \BibitemOpen
  \bibfield  {author} {\bibinfo {author} {\bibfnamefont {S.}~\bibnamefont
  {Mertens}},\ }\href {\doibase 10.1007/BF01026565} {\bibfield  {journal}
  {\bibinfo  {journal} {J. Stat. Phys.}\ }\textbf {\bibinfo {volume} {58}},\
  \bibinfo {pages} {1095} (\bibinfo {year} {1990})}\BibitemShut {NoStop}%
\bibitem [{lpd()}]{lpdata}%
  \BibitemOpen
  \href@noop {} {\enquote {\bibinfo {title} {Duke digital repository},}\
  }\bibinfo {howpublished} {https://doi.org/10.7924/xxxxxxxxx}\BibitemShut
  {NoStop}%
\bibitem [{\citenamefont {Wilkinson}\ and\ \citenamefont
  {Willemsen}(1983)}]{wilkinson1983invasion}%
  \BibitemOpen
  \bibfield  {author} {\bibinfo {author} {\bibfnamefont {D.}~\bibnamefont
  {Wilkinson}}\ and\ \bibinfo {author} {\bibfnamefont {J.~F.}\ \bibnamefont
  {Willemsen}},\ }\href {\doibase 10.1088/0305-4470/16/14/028} {\bibfield
  {journal} {\bibinfo  {journal} {J. Phys. A}\ }\textbf {\bibinfo {volume}
  {16}},\ \bibinfo {pages} {3365} (\bibinfo {year} {1983})}\BibitemShut
  {NoStop}%
\bibitem [{\citenamefont {Van~der Marck}(1998)}]{van1998calculation}%
  \BibitemOpen
  \bibfield  {author} {\bibinfo {author} {\bibfnamefont {S.~C.}\ \bibnamefont
  {Van~der Marck}},\ }\href {\doibase 10.1142/S0129183198000431} {\bibfield
  {journal} {\bibinfo  {journal} {Int. J. Mod. Phys. C}\ }\textbf {\bibinfo
  {volume} {9}},\ \bibinfo {pages} {529} (\bibinfo {year} {1998})}\BibitemShut
  {NoStop}%
\bibitem [{\citenamefont {Gaunt}\ \emph {et~al.}(1976)\citenamefont {Gaunt},
  \citenamefont {Sykes},\ and\ \citenamefont {Ruskin}}]{gaunt1976percolation}%
  \BibitemOpen
  \bibfield  {author} {\bibinfo {author} {\bibfnamefont {D.}~\bibnamefont
  {Gaunt}}, \bibinfo {author} {\bibfnamefont {M.}~\bibnamefont {Sykes}}, \ and\
  \bibinfo {author} {\bibfnamefont {H.}~\bibnamefont {Ruskin}},\ }\href
  {\doibase 10.1088/0305-4470/9/11/015} {\bibfield  {journal} {\bibinfo
  {journal} {J. Phys. A}\ }\textbf {\bibinfo {volume} {9}},\ \bibinfo {pages}
  {1899} (\bibinfo {year} {1976})}\BibitemShut {NoStop}%
\bibitem [{\citenamefont {Gaunt}\ and\ \citenamefont
  {Ruskin}(1978)}]{gaunt1978bond}%
  \BibitemOpen
  \bibfield  {author} {\bibinfo {author} {\bibfnamefont {D.}~\bibnamefont
  {Gaunt}}\ and\ \bibinfo {author} {\bibfnamefont {H.}~\bibnamefont {Ruskin}},\
  }\href {\doibase 10.1088/0305-4470/11/7/025} {\bibfield  {journal} {\bibinfo
  {journal} {J. Phys. A}\ }\textbf {\bibinfo {volume} {11}},\ \bibinfo {pages}
  {1369} (\bibinfo {year} {1978})}\BibitemShut {NoStop}%
\bibitem [{\citenamefont {Pordes}\ \emph {et~al.}(2007)\citenamefont {Pordes},
  \citenamefont {Petravick}, \citenamefont {Kramer}, \citenamefont {Olson},
  \citenamefont {Livny}, \citenamefont {Roy}, \citenamefont {Avery},
  \citenamefont {Blackburn}, \citenamefont {Wenaus}, \citenamefont
  {W{\"u}rthwein}, \citenamefont {Foster}, \citenamefont {Gardner},
  \citenamefont {Wilde}, \citenamefont {Blatecky}, \citenamefont {McGee},\ and\
  \citenamefont {Quick}}]{osg07}%
  \BibitemOpen
  \bibfield  {author} {\bibinfo {author} {\bibfnamefont {R.}~\bibnamefont
  {Pordes}}, \bibinfo {author} {\bibfnamefont {D.}~\bibnamefont {Petravick}},
  \bibinfo {author} {\bibfnamefont {B.}~\bibnamefont {Kramer}}, \bibinfo
  {author} {\bibfnamefont {D.}~\bibnamefont {Olson}}, \bibinfo {author}
  {\bibfnamefont {M.}~\bibnamefont {Livny}}, \bibinfo {author} {\bibfnamefont
  {A.}~\bibnamefont {Roy}}, \bibinfo {author} {\bibfnamefont {P.}~\bibnamefont
  {Avery}}, \bibinfo {author} {\bibfnamefont {K.}~\bibnamefont {Blackburn}},
  \bibinfo {author} {\bibfnamefont {T.}~\bibnamefont {Wenaus}}, \bibinfo
  {author} {\bibfnamefont {F.}~\bibnamefont {W{\"u}rthwein}}, \bibinfo {author}
  {\bibfnamefont {I.}~\bibnamefont {Foster}}, \bibinfo {author} {\bibfnamefont
  {R.}~\bibnamefont {Gardner}}, \bibinfo {author} {\bibfnamefont
  {M.}~\bibnamefont {Wilde}}, \bibinfo {author} {\bibfnamefont
  {A.}~\bibnamefont {Blatecky}}, \bibinfo {author} {\bibfnamefont
  {J.}~\bibnamefont {McGee}}, \ and\ \bibinfo {author} {\bibfnamefont
  {R.}~\bibnamefont {Quick}},\ }in\ \href {\doibase
  10.1088/1742-6596/78/1/012057} {\emph {\bibinfo {booktitle} {J. Phys. Conf.
  Ser.}}},\ \bibinfo {series} {78}, Vol.~\bibinfo {volume} {78}\ (\bibinfo
  {year} {2007})\ p.\ \bibinfo {pages} {012057}\BibitemShut {NoStop}%
\bibitem [{\citenamefont {Sfiligoi}\ \emph {et~al.}(2009)\citenamefont
  {Sfiligoi}, \citenamefont {Bradley}, \citenamefont {Holzman}, \citenamefont
  {Mhashilkar}, \citenamefont {Padhi},\ and\ \citenamefont
  {Wurthwein}}]{osg09}%
  \BibitemOpen
  \bibfield  {author} {\bibinfo {author} {\bibfnamefont {I.}~\bibnamefont
  {Sfiligoi}}, \bibinfo {author} {\bibfnamefont {D.~C.}\ \bibnamefont
  {Bradley}}, \bibinfo {author} {\bibfnamefont {B.}~\bibnamefont {Holzman}},
  \bibinfo {author} {\bibfnamefont {P.}~\bibnamefont {Mhashilkar}}, \bibinfo
  {author} {\bibfnamefont {S.}~\bibnamefont {Padhi}}, \ and\ \bibinfo {author}
  {\bibfnamefont {F.}~\bibnamefont {Wurthwein}},\ }in\ \href {\doibase
  10.1109/CSIE.2009.950} {\emph {\bibinfo {booktitle} {2009 WRI World Congress
  on Computer Science and Information Engineering}}},\ \bibinfo {series} {2},
  Vol.~\bibinfo {volume} {2}\ (\bibinfo {year} {2009})\ pp.\ \bibinfo {pages}
  {428--432}\BibitemShut {NoStop}%
\bibitem [{\citenamefont {Convay}\ and\ \citenamefont
  {Sloane}(1982)}]{convay1982fast}%
  \BibitemOpen
  \bibfield  {author} {\bibinfo {author} {\bibfnamefont {J.~H.}\ \bibnamefont
  {Convay}}\ and\ \bibinfo {author} {\bibfnamefont {N.~J.~A.}\ \bibnamefont
  {Sloane}},\ }\href {\doibase 10.1109/TIT.1982.1056484} {\bibfield  {journal}
  {\bibinfo  {journal} {IEEE Trans. Inf. Theory}\ }\textbf {\bibinfo {volume}
  {28}},\ \bibinfo {pages} {227} (\bibinfo {year} {1982})}\BibitemShut
  {NoStop}%
\bibitem [{\citenamefont {Conway}\ and\ \citenamefont
  {Sloane}(1995)}]{conway1995all}%
  \BibitemOpen
  \bibfield  {author} {\bibinfo {author} {\bibfnamefont {J.~H.}\ \bibnamefont
  {Conway}}\ and\ \bibinfo {author} {\bibfnamefont {N.~J.~A.}\ \bibnamefont
  {Sloane}},\ }\href {\doibase 10.1007/BF02574051} {\bibfield  {journal}
  {\bibinfo  {journal} {Discrete Comput. Geom.}\ }\textbf {\bibinfo {volume}
  {13}},\ \bibinfo {pages} {383} (\bibinfo {year} {1995})}\BibitemShut
  {NoStop}%
\bibitem [{\citenamefont {Lorenz}\ \emph {et~al.}(2000)\citenamefont {Lorenz},
  \citenamefont {May},\ and\ \citenamefont {Ziff}}]{lorenz2000similarity}%
  \BibitemOpen
  \bibfield  {author} {\bibinfo {author} {\bibfnamefont {C.~D.}\ \bibnamefont
  {Lorenz}}, \bibinfo {author} {\bibfnamefont {R.}~\bibnamefont {May}}, \ and\
  \bibinfo {author} {\bibfnamefont {R.~M.}\ \bibnamefont {Ziff}},\ }\href
  {\doibase 10.1023/A:1018648130343} {\bibfield  {journal} {\bibinfo  {journal}
  {J. Stat. Phys.}\ }\textbf {\bibinfo {volume} {98}},\ \bibinfo {pages} {961}
  (\bibinfo {year} {2000})}\BibitemShut {NoStop}%
\end{thebibliography}%

\end{document}